\documentclass[prd,preprintnumbers,twocolumn,nofootinbib,showpacs]{revtex4}

\usepackage{amsfonts,amsmath,amssymb}
\usepackage{bm}
\usepackage{graphicx,epsfig}
\usepackage{dcolumn}

\newcommand{\gsim}{\mathrel{\lower2.5pt\vbox{\lineskip=0pt\baselineskip=0pt
                   \hbox{$>$}\hbox{$\sim$}}}}
\newcommand{\lsim}{\mathrel{\lower2.5pt\vbox{\lineskip=0pt\baselineskip=0pt
                   \hbox{$<$}\hbox{$\sim$}}}}

\newcommand{\too}{\longrightarrow}

\newcommand{\abs}[1]{\left| #1 \right|}
\newcommand{\avg}[1]{\left\langle #1 \right\rangle}

\newcommand{\sla}[1]{{\raise.15ex\hbox{$/$}\kern-.57em #1}}
\newcommand{\Sla}[1]{\kern0.12em{\raise.15ex\hbox{$/$}\kern-.74em #1}}

\newcommand{\ev}{{\rm eV}}

\newcommand{\Mev}{{\rm MeV}}
\newcommand{\gev}{{\rm GeV}}
\newcommand{\tev}{{\rm TeV}}
\newcommand{\Mpl}{M_{\rm P\ell}}

\newcommand{\mW}{m_{\text{{\tiny $W$}}}}

\newcommand{\beq}{\begin{eqnarray}}
\newcommand{\eeq}{\end{eqnarray}}
\newcommand{\nn}{\nonumber}
\newcommand{\eql}[1]{\label{eq:#1}}
\newcommand{\eq}[1]{(\ref{eq:#1})}

\newcommand{\colmvec}[1]{\left( \begin{array}{c} #1 \end{array} \right)}
\newcommand{\order}[1]{{\mathcal O}\!\left( #1 \right)}

\newcommand{\ie}{\textit{i.e.}}

\newcommand{\Lg}{\Lambda_g}
\newcommand{\Lgt}{\widetilde{\Lambda}_g}
\newcommand{\chit}{{\widetilde{\chi}}}
\newcommand{\Xt}{{\widetilde{X}}}
\newcommand{\micr}{{\rm \mu m}}

\begin{document}

\preprint{BUHEP-05-15}
\preprint{hep-ph/0511082}

\title{Probing Composite Gravity in Colliders}

\author{Takemichi Okui}
\affiliation{Physics Department, Boston University,\\
             590 Commonwealth Ave., Boston, MA 02215, U.S.A.}


\begin{abstract}
We explore scenarios in which the graviton is not a fundamental degree of 
freedom at short distances but merely emerges as an effective 
degree of freedom at long distances.  In general, the scale of such 
graviton `compositeness', $\Lg$, can only be probed by measuring gravitational 
forces at short distances, which becomes increasingly difficult and eventually
impossible as the distance is reduced.  Here, however, we point out that if 
supersymmetry is an underlying symmetry, the gravitino can be used as an 
alternative probe to place a limit on $\Lg$ in a collider environment, 
by demonstrating that there is a model-independent relation, 
$\Lg \gsim m_{3/2}$.  In other 
words, the gravitino knows that gravity is standard at least down to its 
Compton wavelength, so this can also be viewed as a test of general 
relativity possible at very short distances.  If composite gravity is found 
first at some $\Lg$, this would imply a model-independent upper bound on 
$m_{3/2}$.        
\end{abstract}

\maketitle


\section{Introduction}
\label{sec:intro}
Gravity at short distances is a vastly unexplored experimental frontier.
It is possible that a deviation or even a drastic departure from the 
standard gravitational law may be found in future experiments.  
On the theoretical side, we have string theory which replaces general 
relativity (GR) at distances shorter than the string scale $M_s^{-1}$.  
However, since string theory not only modifies gravity but also governs the 
matter sector, the fact that we have not observed any stringy phenomena in 
particle physics experiments requires $M_s$ to be higher than at least a 
few $\tev$.

In contrast, for a theory which modifies \textit{only} gravity, 
the bound on the scale of such new short-distance gravitational physics is 
significantly lowered to 
$(\order{100}\,\micr)^{-1} \approx \order{10^{-3}}\,\ev$ 
\cite{Hoyle:2004cw} \cite{Long:2003dx, Smullin:2005iv} (also see the review 
\cite{Adelberger:2003zx}), which is 15 orders of magnitude larger than 
$\tev^{-1} \approx 10^{-17}\,{\rm cm}$!  Therefore, there is {\it huge}
room for a theory of this kind.  This situation is quite intriguing, and this 
is the window that we will explore in this paper.
  
The striking fact about this range 
between $100\,\micr$ and $10^{-17}\,{\rm cm}$
is that we {\it know} that matter is 
described by the standard local relativistic quantum field theory there.  
The standard model (SM) has been tested including nontrivial loop
corrections with great precision \cite{:2005em}.  This point cannot be 
emphasized too much.  It means that a modification of gravity in this range
cannot be as radical as, for example, abandoning the notion of a continuum 
spacetime; when we say the Bohr radius is $0.509$~\AA, we know perfectly 
what we are talking about!  So, while we will boldly speak of modifying 
gravity in this paper, \textit{we will not mess around with matter;} 
we take it for granted that the matter sector is completely 
\textit{normal,} \ie, perfectly described by a local relativistic 
quantum field theory.

It should be also mentioned that, in general, changing the laws of gravity 
does not necessarily mean modifying or abandoning GR.  For example, if we 
add $n$ extra spatial dimensions with the size $L$ in which only gravitons may 
propagate, then the Newton's law changes from $1/r^2$ to $1/r^{2+n}$ for 
$r \ll L$ \cite{Arkani-Hamed:1998rs}.  But gravity in this example is 
perfectly governed by the conventional GR; it is just living in more 
dimensions than four.  

In this paper, however, we \textit{will} explore the possibility that GR 
is abandoned at short distances in the sense that the graviton is \textit{not} 
a fundamental propagating degree of freedom (d.o.f.)~in whatever underlying 
theory, but is merely an effective d.o.f.~appropriate at long distances.  
The scale, which we call $\Lg^{-1}$, corresponding to the boundary between 
`short-distances' and `long-distances' could be anywhere shorter than 
$\order{100}\,\micr$, but as we stated above, we will focus on the range
$10^{-17}\,{\rm cm} \lsim \Lg^{-1} \lsim 100\,\micr$ (or 
$10^{-3}\>\ev \lsim \Lg \lsim \tev$), so that we can exploit the fact that 
the matter sector is `normal'.

This includes various possibilities---the graviton may be a bound 
or solitonic state of the fundamental d.o.f.~\cite{Sundrum:1997js}, or an 
extended state in some intrinsically nonlocal theory 
\cite{Moffat:2001jf, Sundrum:2003tb}, or a sort of hydrodynamic state as in 
the scenarios often dubbed `emergent relativity' \cite{Unruh:1980cg}.
We will not distinguish these varieties but just focus on their common 
feature that the graviton is not an elementary propagating d.o.f.~in the 
fundamental theory but just appears as an effective d.o.f.~in the long-distance 
description for $d>\Lg^{-1}$.  Admittedly not an optimal name, we call it a 
{\it composite} graviton, {\it where by `composite' we simply mean `not 
elementary'.} 

\begin{figure*}
\includegraphics{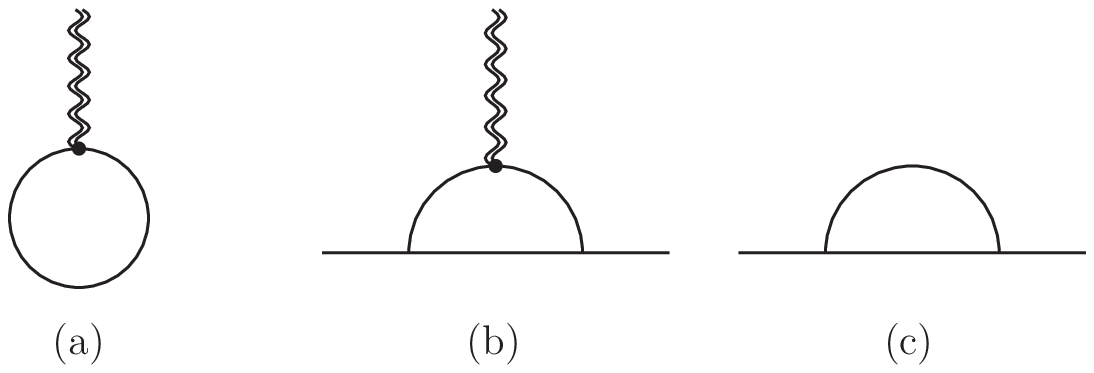}
\caption{Diagrams representing (a) corrections to the cosmological constant,
(b) corrections to the gravitational mass, and (c) corrections to the inertial
mass.  A solid line represents a heavy matter particle, and a double wavy 
line represents a graviton.}
\label{fig:diagrams}
\end{figure*}

One may think such a composite graviton is excluded by the theorem by 
Weinberg and Witten \cite{Weinberg:1980kq}.  Actually, what the 
Weinberg-Witten (WW) theorem excludes is not just a composite graviton but 
\textit{any} massless spin-1 or -2 particle, {\it composite or not!}  
Therefore, we must be careful about the assumptions of the theorem; we all 
know QED and QCD which have a massless spin-1 particle, and GR which has a 
massless spin-2 particle.  Note that the WW theorem 
states that if a theory allows the existence of a Lorentz-covariant conserved 
vector (or symmetric 2nd-rank tensor) current, then the theory cannot contain 
any massless spin-1 (or spin-2) particle charged under this current.  QED 
evades the spin-1 part of the theorem because the photon is not charged under 
the current.  QCD evades the spin-1 part of the theorem because the current is 
not Lorentz covariant due to its dependence on the gluon field which is a 
4-vector only up to a gauge transformation.  Similarly, GR evades the spin-2 
part of the theorem because the gravitational part of the energy-momentum 
`tensor' is not really a tensor in GR. 

Indeed, there is an {\it explicit} example of composite gauge bosons.  
Consider an $SU(N)$ supersymmetric QCD with $F$ flavors where 
$N+1 < F < 3N/2$, with no superpotentials.  In the far infrared (IR), this 
theory is described by an IR-free, weakly coupled $SU(F-N)$ gauge theory 
\cite{Seiberg:1994pq}.  However, these IR gauge bosons 
are {\it not}  a subset of the original ultraviolet (UV) d.o.f.; rather, 
they are new effective d.o.f.~appearing only in the IR description, which 
microscopically can be interpreted as solitonic states of the fundamental UV 
d.o.f.~\cite{Seiberg:1994pq}.  So this is indeed a \textit{concrete} example 
of composite massless gauge bosons, where the $SU(F-N)$ gauge symmetry emerges
at low energies, making it consistent with the WW theorem.  

Clearly, it is desirable to have a similar example for gravity.  To this 
goal, Gherghetta, Peloso and Poppitz recently presented a theory in a 
5-dimensional Anti-de-Sitter (AdS) space which is dual to a 4-dimensional 
conformal field theory in which the conformal symmetry is dynamically broken
in the IR yielding a spectrum containing a massless spin-2 resonance 
\cite{Gherghetta:2005se}.  To complete their picture, analyses beyond the 
quadratic order in action must be performed, especially concerning effects 
of the stabilization of the AdS space, and if the delicate existence of the 
massless spin-2 state persists, this will be a solid, concrete example of 
a theory of a composite graviton.  (Note that the graviton in string theory 
is completely elementary.) 

Is there any reason or motivation to consider such drastic modification of 
gravity in this range?  Just near the edge of the range, there is a 
cosmologically interesting scale 
$\approx (20\,\micr)^{-1} \approx 10^{-2}\>\ev 
\approx (16\pi^2\rho_{\rm vac})^{1/4}$, where 
$\rho_{\rm vac}$ is the vacuum energy density corresponding to the observed 
acceleration of the expansion of the universe 
\cite{Perlmutter:1998np, Spergel:2003cb, Tegmark:2003ud}.  Kaplan and Sundrum 
also recently pointed out that the interesting scale in the context of the
cosmological constant problem (CCP) may instead be $\order{10}\,\Mev$ 
\cite{Kaplan:2005rr}.  
Therefore, it is quite interesting to ask if composite gravity can solve the
CCP by identifying $\Lg$ with, say, $10^{-2}\>\ev$.
However, it is not so hard to see the answer entirely depends on the nature of 
whatever underlying theory of composite gravity.

In particular, it appears that the underlying theory should not be a local
field theory if one wishes to suppress loop corrections to the cosmological 
constant by abandoning elementary gravitons \cite{Sundrum:2003jq}.  The 
argument goes as follows.  Consider three diagrams in FIG.~\ref{fig:diagrams}.
The diagram (\ref{fig:diagrams}-a) is a correction to the vacuum energy, 
(\ref{fig:diagrams}-b) is a correction to the gravitational mass, and 
(\ref{fig:diagrams}-c) is a correction to the inertial mass.  In a field 
theory, the loop integral in (\ref{fig:diagrams}-a) can be suppressed only if 
the vertex has a form factor that depends on the loop momentum.  Now, the 
problem is, once (\ref{fig:diagrams}-a) is suppressed by such a form factor, 
the correction (\ref{fig:diagrams}-b) also gets suppressed because it has the 
same form factor, while the correction (\ref{fig:diagrams}-c) does not get 
suppressed because there is no such form factor.  This violates the 
equivalence principle, and we need fine-tuning to restore it.  However, for a 
composite graviton which is not from a local field theory, there does not have
to be tension like this, 
and suppressing loop corrections to the cosmological constant may 
be consistent with the equivalence principle.  But even supposing we did
find such a nonlocal underlying theory, it would still be 
halfway to solving the cosmological constant problem, since there 
are also {\it tree-level} or classical contributions to the vacuum energy from 
phase transitions which must be somehow suppressed.  The door is not shut yet, 
and Ref.~\cite{Sundrum:2003tb} discusses a toy model for such a nonlocal theory 
without problems with the equivalence principle or the classical contributions.  
In the rest of the paper, we will not concern ourselves with the cosmological 
constant problem any further, and just focus on the physics of composite 
gravity. 

So, supposing that the graviton is not an elementary d.o.f.~in whatever 
fundamental theory, how do we see it?  Without having a concrete microscopic 
model of composite gravity, the scale $\Lg$ is the only quantity we can 
discuss.  So far, the lower bound on $\Lg$ has been placed by measuring 
gravitational forces between test masses, which has reached the scale of 
$\order{100}\,\micr$.  But it is clear that such direct measurement will be 
increasingly difficult and eventually impossible as the distance gets reduced.
Soon, some other methods must replace it to probe the scale of composite 
gravity.  

Such an alternative can arise if there is something that is related to the 
graviton 
but is more accessible than the graviton at short distances.  In general, 
there is nothing that is related to gravity except the graviton itself.  
However, if nature possesses (spontaneously broken) supersymmetry (SUSY), the 
{\it gravitino} precisely satisfies the criteria---it is related to the 
graviton and may be accessible even in colliders!  
The introduction of SUSY allows us to extract some informations 
relating the graviton and the gravitino without knowing what the underlying
theory is.  In fact, we will show that if a gravitino 
exists, it can indeed be used to probe gravity at very short distances where 
direct measurement of gravitational forces is impossible. 

To keep our discussions as model independent as possible, we would like to
have an effective field theory and ask questions that can be answered by
it.  This effective theory must have the following features:
\begin{itemize}
 \item{It must contain a {\it physical} scale $\Lg$ above which the graviton is
       no longer an elementary degree of freedom.  The scale $\Lg$ is not a 
       scale chosen for convenience but corresponds to a physical boundary 
       between two completely different phases of the theory, just like 
       $\Lambda_{\rm QCD}$ separates two different descriptions with totally 
       different degrees of freedom (\ie~partons versus hadrons).}
 \item[]{(Recall that $\Lg$ is a parameter anywhere from $\order{10^{-3}}\,\ev$ 
       to $\order{\tev}$ or whatever cutoff for the matter sector.)}
 \item{Nevertheless, to reproduce all the known gravitational physics, it 
       must include all the matter particles, \textit{even the ones heavier than 
       $\Lg$!}  And, as emphasized already, we know that the 
       matter sector is perfectly described by a local relativistic quantum 
       field theory with a cutoff higher than $\tev > \Lg$.}
\end{itemize}
Because of the second feature, we cannot use the usual effective field theory 
formalism in which all the particles heavier than $\Lg$ are simply integrated 
out; that would fail to capture all the {\it known} long-distance gravitational 
physics such as the $1/r^2$ law, the perihelion precession, the bending of 
light, etc.

Therefore, the first important question is whether or not there exists a 
sensible effective theory that can deal with this highly asymmetric situation 
in which gravity has a low cutoff and matter has a high cutoff.  This question 
was answered by R.~Sundrum, who developed a formalism, {\it soft graviton 
effective theory} (SGET) \cite{Sundrum:2003jq}, which assures that we can 
consistently analyze this asymmetric situation without referring to the 
underlying theory of composite gravity.  We will review the essential ideas 
of SGET in Sec.~\ref{sec:SGET} to keep our discussions self-contained.

Given that there is a consistent effective field theory to describe the 
low-cutoff gravity with the high-cutoff heavy matter, there seems nothing 
wrong to have a gravitino heavier than $\Lg$, since we should be able to 
treat it just as one of heavy matter particles.  After all, $\Lg$ is the 
scale of {\it graviton's} compositeness which does not have to be equal to
that of {\it gravitino's} once supersymmetry is broken.  Also, there is 
nothing wrong {\it a priori} for a composite particle to be heavier than the 
scale of its compositeness, like the $B$-mesons, the hydrogen atom, etc.  

Nevertheless, as we will show in Sec.~\ref{sec:SusySGET}, there is a nontrivial
remnant of the underlying supersymmetry which gives rise to the relation
\beq
   m_{3/2} \lsim \Lg ~.
\eeq 
Therefore, in fact a gravitino---if it exists---{\it knows} that gravity 
should be just GR (\ie~the graviton is an elementary d.o.f.)~at least down to 
its Compton wavelength!  In other words, the discovery of a gravitino and the 
measurement of its mass offers a short-distance test of GR and places a 
{\it model-independent} lower-bound on $\Lg$!  In particular, depending on the 
value of $m_{3/2}$, we may be able to completely exclude the possibility of 
composite gravity as a solution to the CCP.  

On the other hand, if we first 
discover composite gravity somehow and measure $\Lg$ before discovering a 
gravitino, then this inequality predicts that, once we see a gravitino, we 
will find its mass be lighter than $\Lg$.  

In Sec.~\ref{sec:existence} and \ref{sec:relation}, we will continue the
discussions to gain a further understanding of the inequality, followed 
by a brief comment in Sec.~\ref{sec:theo-tests} on the possibility of 
independent theoretical tests of the inequality. 

In order for our prediction to be useful, it is clearly crucial to 
experimentally convince ourselves that what we are observing is really a 
gravitino, not a random spin-$3/2$ resonance which may just happen to be 
there.  This issue will be discussed in Sec.~\ref{sec:detection}.  We will 
then conclude in Sec.~\ref{sec:conc}.

\section{Soft Graviton Effective Theory}
\label{sec:SGET}
As we have already mentioned, we need to describe all 
experimentally known gravitational physics occurring among heavy ($\gg \Lg$) 
matter particles, without extrapolating our knowledge of gravity beyond $\Lg$.
Soft graviton effective theory (SGET) \cite{Sundrum:2003jq} is designed 
precisely for this purpose.%
\footnote{Strictly speaking, to describe the typical observed gravitational 
phenomena involving gravitational bound states, we should switch to yet another 
effective field theory to have a transparent power-counting 
scheme appropriate for that purpose.  The interested reader should read 
Ref.~\cite{Goldberger:2004jt} which develops such an effective theory, 
dubbed `nonrelativistic general relativity' (NRGR).}
Here, we will review its central concepts to keep the discussions 
self-contained.
  
To start, let us consider gravity only.  In this case, the theory takes the 
form of a familiar effective field theory with the cutoff $\Lg$ imposed on 
the graviton field $h_{\mu\nu}$ defined via
\beq
   g_{\mu\nu} = \eta_{\mu\nu} + \frac{h_{\mu\nu}}{\Mpl} ~.
\eeq
Namely, the lagrangian is just the usual Ricci scalar term plus 
a whole series of higher-dimensional operators suppressed by powers of $\Lg$:
\beq
   {\cal L}_{\rm grav} 
      \sim \Mpl^2 \left( {\cal R} 
                         + \frac{{\cal R}^2}{\Lg^2}
                         + \frac{R_{\mu\nu} R^{\mu\nu}}{\Lg^2}
                         + \cdots
                         \right)  ~,
\eql{puregrav}
\eeq
where dimensionless $\order{1}$ coefficients are suppressed.%
\footnote{The operator $R_{\mu\nu\rho\sigma} R^{\mu\nu\rho\sigma}$ can be 
omitted in perturbation theory since it can be expressed as a linear 
combination of the two operators explicitly written in \eq{puregrav} plus
a total derivative.  Furthermore, in the absence of matter, even those two
operators could be removed by field redefinition, but we have kept 
them in \eq{puregrav} because we are interested in including matter.  
See Refs.~\cite{Donoghue:1995cz} for more discussions on these operators.}
As we mentioned earlier, $\Lg$ is a physical scale above which $h_{\mu\nu}$ 
is no longer an elementary degree of freedom.    

Note the scales and kinematic configurations to which this 
${\cal L}_{\rm grav}$ is 
applicable.  It is appropriate only for processes where all momentum 
transfers among the gravitons are less than $\order{\Lg}$.  For example, it 
can {\it not} be used to calculate the cross-section 
for two highly energetic ($E \gg \Lg$) gravitons scattering with a large 
angle.  In fact, we do not even know if such a scattering occurs at 
all---maybe they would end up with `jets', like in hadron-hadron 
collisions---who knows?  No experiments so far have told us what would happen to 
such processes, and performing theoretical calculations requires specifying 
the full theory valid at distances shorter than $\Lg^{-1}$.  The moral here
is that a large momentum transfer should not be delivered to a 
graviton within our effective theory.

To also understand that a graviton should not be {\it exchanged} 
to mediate a large momentum transfer, imagine a theory with a fermion 
$\psi$ and a scalar $\phi$, and suppose we have verified that a Yukawa coupling 
$\phi\bar{\psi}\psi$ perfectly describes the $\psi\psi \to \psi\psi$ scattering
when both $\psi$'s get only very {\it low} recoils, \ie~the momentum transfer 
mediated by $\phi$ is very small.  But it may be completely wrong to use this 
Yukawa theory to describe the scattering of two very energetic $\psi$'s by a 
large angle, corresponding to a large momentum transfer mediated by $\phi$.  
For instance, suppose that the $\phi$ is actually a strongly bound state of 
two {\it new} fermions $\Psi$ interacting with $\psi$ via a 4-fermion coupling 
$\bar{\psi}\psi\,\overline{\Psi}\Psi$.  Then, when the $\psi$'s get recoils 
much larger than `$\Lambda_{\rm QCD}$' of this new strong interaction, we must 
use the 4-fermion theory with $\Psi$ rather than the Yukawa theory with 
$\phi$.  Here, $\phi$ is meant to be the analog of the graviton, and 
therefore, within our effective theory, a graviton should not be exchanged 
to mediate a large momentum transfer ($\gg \Lg$). 

Now, let us move on and include matter fields.  First, note that for a given
value of $\Lg$, some elementary particles in the standard model (SM) are too 
short-lived ($\tau \ll \Lg^{-1}$) to be included in SGET.  On the other hand, 
some composite particles in the SM live long enough ($\tau \gg \Lg^{-1}$) and 
also are too small in size ($\ll \Lg^{-1}$) for a soft graviton to recognize
that they are composite.  For example, if $\Lg$ is, say, $10^{-2}\>\ev$, then 
the proton, the hydrogen atoms in $1S$ and $2P$ states would be all 
{\it elementary} fields (fermion, scalar, and vector, respectively) in SGET.  

Secondly, there are many `hard processes' among those matter particles involving 
momentum transfers much larger than $\Lg$.  For example, if $\Lg$ is, say, 
$1\>\ev$, then the pair annihilation, $e^+ e^- \to \gamma \gamma$, would be a 
hard process.  Since a soft graviton in this case cannot resolve the $t$-channel 
electron propagator there, we should shrink it to a point and express the entire 
process by a {\it single} local operator.  Also, since soft gravitons in this
case cannot pair-produce an electron and a positron, they are completely 
unrelated particles from soft gravitons' viewpoint.  Whereas if $\Lg$ is, say,
$1\>\gev$, then there are soft gravitons who can see the $t$-channel propagator 
in $e^+ e^- \to \gamma\gamma$, and electrons and positrons must be described by 
a single field operator.

The general matching procedure for a SGET may be best explained by comparing 
it with the construction of a usual effective field 
theory in which heavy particles are simply integrated out.  In the derivation 
of a usual effective theory, we consider 
\textit{one-light-particle-irreducible} (1LPI) diagrams; in a 1LPI diagram, 
all external lines represent light particles to be kept in the effective 
theory, and the diagram would not split in two if any one of internal 
light-particle propagators were cut.  We then obtain effective vertices in the 
effective theory by shrinking every heavy propagator to a point.

Similarly, for a SGET, we consider 
\textit{one-nearly-on-shell-particle-irreducible} (1NOSPI) diagrams; in a 
1NOSPI diagram, all external lines are nearly on-shell, \ie, its deviation 
from the mass shell is less than $\order{\Lg}$, and the diagram would not
split in two if any one of internal nearly-on-shell propagators were cut.  We 
then obtain effective vertices by shrinking every far-off-shell (\ie~not nearly 
on-shell) propagator to a point.  For the technical detail of `shrinking' 
or matching procedure, see Ref.~\cite{Sundrum:2003jq}.

Having matched all hard SM processes onto effective operators, 
we are now ready to couple to it the soft graviton described by 
\eq{puregrav}.  This step is straightforward---we just 
use general covariance as a guide, just as we do for the conventional general 
relativity.

By construction, SGET respects all fundamental requirements such as 
the equivalence principle, Lorentz invariance, and unitarity, as long as we 
stay within its applicability we have discussed above \cite{Sundrum:2003jq}.  
In particular, unitarity 
holds because all propagators that can be on-shell are included in SGET, so it 
correctly reproduces the imaginary part of any amplitude.

\section{The Composite Gravitino}
\label{sec:SusySGET}
Now, let us consider putting a gravitino in the story, with the hope that
a gravitino may be more experimentally accessible than gravitons at short 
distances so we can learn something about gravity.  The new ingredient in this 
section is supersymmetry (SUSY) as a spontaneously broken exact underlying 
symmetry, not only in the matter sector {\it but also in the gravity sector.}  
As mentioned in Sec.~\ref{sec:intro}, the introduction of SUSY is a necessary 
and minimal additional ingredient if we wish to have an alternative probe for 
$\Lg$ which is as model-independent as possible, because without SUSY there is 
nothing that is necessarily related to gravity except the graviton itself.    

Since the graviton is not a fundamental degree of freedom at
short distances, neither is the gravitino.%
\footnote{Of course, there is also a possibility that a gravitino just does not 
exist.  Here, we assume that a gravitino exists and its lifetime is long
enough ($\gg \Lg^{-1}$) to be in the effective theory.  We will come back to 
this caveat in Sec.~\ref{sec:existence}.}
Let $\Lgt$ be the scale above which the gravitino ceases to be an elementary 
degree of freedom.  Because supersymmetry is broken, $\Lgt$ does not have to 
be equal to $\Lg$.  There is also another scale in the theory, the gravitino 
mass $m_{3/2}$.  {\it A priori, these three scales may come in any order.}  
SGET assures that there is a consistent framework to describe
particles which are much heavier than $\Lg$, so $m_{3/2}$ may be higher or 
lower than $\Lg$.  While $\Lgt$ is roughly the `size' of the gravitino, 
there is nothing wrong for a composite particle to be heavier than the 
inverse of its size, or the compositeness scale.  In fact, heavy quark 
effective theory (HQET) \cite{Georgi:1990um}, which describes a single 
$B$-meson system, takes advantage of the fact that the $B$-meson's 
compositeness scale $\Lambda_{\rm QCD}$ is much less than its mass 
$m_B \approx m_b$.        

In the case of HQET, the effective theory breaks down if a gluon delivers
a momentum transfer larger than $m_b$ to the $b$ quark.  But in general 
effective theories, the breakdown may happen at an energy much lower than 
any obvious mass scale in the theory.  For example, consider the effective 
field theory of a hydrogen atom in the ground state interacting with soft 
photons ($E_\gamma \ll \order{\ev}$).  This effective theory contains an 
elementary scalar field (the hydrogen atom in the $1S$ state) and the 
electromagnetic field, and it correctly accounts for the Rayleigh scattering, 
explaining why the sky is blue.%
\footnote{The reader not familiar with this cute application of
effective field theory may like to read Ref.~\cite{Kaplan:1995uv}.}
But this effective theory clearly goes wrong if a photon delivers an energy 
of $\order{\ev}$ or higher, where we should take into account the
fact that the scalar is actually not elementary.  But this breakdown scale 
is much less than the scalar mass, $\order{\gev}$.%
\footnote{We get a different breakdown scale if we are interested in 
capturing a different physics, such as the pair-annihilation of a hydrogen
and an anti-hydrogen.}%
$^{,}$%
\footnote{Interestingly, even if we take into account the internal structure,
the breakdown scale $\order{\ev}$ is still much smaller than the lightest mass
in the theory $m_e \approx .5~\Mev$.  See Ref.~\cite{Luke:1996hj} for an
illuminating formalism making this breakdown scale manifest.}

Therefore, {\it a priori} there seems no restriction on possible values
for $m_{3/2}$.  (For further discussions shedding different light on this 
matter, see Sec.~\ref{sec:existence}.)  Nevertheless, we will show
below that $m_{3/2}$ should be bounded from above by $\Lg$, which is a 
nontrivial constraint arising from the underlying supersymmetry. 

First, we must be clear about what we mean by `gravitino'.  For instance,
say, we have found a new spin-$3/2$ fermion which has no 
$SU(3)\times SU(2)\times U(1)$ interactions with the rest of the standard
model.  Does it mean we have seen a gravitino?  Not necessarily.  
In order for some spin-$3/2$ fermion to be a candidate for a gravitino, at 
least it must have---possibly among other things---a coupling to the 
supersymmetry current of the matter sector; in other words, it should be 
able to convert a matter particle to its superpartner.  Without this feature, 
it would be no different from a random spin-$3/2$ resonance.

So, we begin by supposing that we have seen a spin-$3/2$ 
fermion $X$ emitted in a process of the type $\widetilde{Y} \to Y + X$, 
where $\widetilde{Y}$ is the superpartner of 
a particle $Y$.  

For $m_{3/2} \ll \Lg$, it is clearly {\it consistent} to add the gravitino in
the pure gravity effective lagrangian \eq{puregrav}, treating it just 
like the graviton.  In other words, we can first forget about the graviton
and gravitino, construct the nearly-on-shell effective lagrangian for matter,
then couple the graviton and the gravitino using general covariance and 
local supersymmetry as a guide, where the effects of $m_{3/2}$ can be 
systematically included as perturbation. 

For $m_{3/2} \gg \Lg$, we clearly cannot include the gravitino in 
\eq{puregrav} together with the soft graviton, because whenever such a heavy 
gravitino is produced or exchanged, it is a hard process ($\gg \Lg$) by 
definition.  
But this simply suggests that we should treat it just as one of heavy matter 
fields instead.  The only difference seems that unlike all the other matter 
particles, we do not have a fundamental theory for the composite gravitino, 
so we cannot calculate the coefficients in SGET lagrangian---that is fine, 
we just leave them as parameters.  

However, we have to be careful, because this splitting of the graviton and 
gravitino into the soft and hard sectors may be incompatible with the 
underlying SUSY, which pairs them.

Let us build a gauge-theory analog of our problem.  First, recall our 
{\it global} symmetry structure: the underlying symmetry is the 
super-Poincar\'e group, which is spontaneously broken to its subgroup, the 
Poincar\'e group.  So, consider a global $SU(2)$ symmetry which spontaneously 
breaks down to a $U(1)$ by a triplet scalar $\phi = \phi^a \sigma^a$ getting 
a VEV $\avg{\phi^{1,2}}=0$ and $\avg{\phi^3}=v$.  (Here, $\phi$ is treated 
just as a spurion.)  The $SU(2)$ is the analog of the underlying supersymmetry, 
while the unbroken $U(1)$ is the analog of the unbroken Poincar\'e symmetry.  

Now, at long distances, the Poincar\'e group is gauged by the existence of the 
soft graviton which, however, is not a fundamental degree of freedom at 
short distances.  So, correspondingly, we gauge the $U(1)$ at long 
distances by introducing a soft massless vector field $W^3_\mu$, which we call 
`toy soft graviton'.  And just like the graviton, $W^3_\mu$ is not a 
propagating degree of freedom at short distances.  Finally, we also need a 
`toy gravitino', \ie, a massive vector 
$W^+_\mu \equiv (W^1_\mu - i W^2_\mu)/\sqrt{2}$.

Let us assume $\mW \gg \Lg$ which is the case of our interest. 
We want to write down `toy SGET' for the toy gravitino.  The only property 
of $W^+_\mu$ which possibly makes it different from other heavy 
particles is that it is the $SU(2)$-partner of the toy graviton $W^3_\mu$.
So, the question is whether there is any constraint on the structure of
the toy SGET from the underlying $SU(2)$, or the toy SUSY.

Let us forget $\Lg$ for a moment, and recall how a spontaneously broken 
symmetry leaves its trace in low-energy physics.  
To be concrete, consider couplings of $W^+_\mu$ and $W^3_\mu$ to a heavy 
Dirac fermion doublet $\psi$.  
($\psi$ is of course the analog of the pair of a SM particle 
and its superpartner.)  If we limit to
only renormalizable operators, all three $W^a_\mu$ must couple to the three 
currents $J^{a\mu} = \bar{\psi}\sigma^a\gamma^\mu\psi$ with a 
single common coupling constant $g$.  This equality is a consequence of the 
underlying $SU(2)$, even though it is broken.  

However, once we take into account higher-dimensional operators, the coupling 
of $W^{1,2}_\mu$ to $J^{1,2}_\mu$ does not have to be equal to that of 
$W^3_\mu$ to $J^3_\mu$, because there are higher dimensional operators that 
reduce to these couplings after picking up the VEV.  Among such, the one with
the lowest dimension is the dimension-5 operator 
$\bar{\psi} \phi \,\Sla{D} \psi$.
We could go on and 
analyze this operator, but it turns out that we can learn the same lesson
with much less arithmetic from the following dimension-6 operator:
\beq
   {\cal L}_6 = -\frac{16\pi^2 c}{M^2} \bar{\psi} \phi \,i\Sla{D} \phi\psi ~,
\eql{dim-6op}
\eeq
where we take $c \sim 1$ so that $M$ corresponds to the scale obtained via
`naive dimensional analysis' (NDA), \ie,
the scale at which this operator would lead to strong coupling if the theory 
is not replaced with a more fundamental theory by then \cite{Manohar:1983md}.  
After substituting the VEV for $\phi$ and canonically normalizing the fields, 
we find that the coupling of $W^3_\mu$ stays equal to $g$ as expected from 
the unbroken $U(1)$ gauge invariance, but the coupling of $W^+_\mu$ does get 
modified as
\beq
   g \too g_+ &=& \frac{1+a}{1-a} \, g  ~,
\eeq
where 
\beq
   a \equiv \frac{16\pi^2 v^2 c}{M^2} \sim \left( \frac{4\pi v}{M} \right)^2 
   ~.
\eeq
Therefore, the equality of the $W^3_\mu$ and $W^+_\mu$ couplings no longer
holds.  Especially, if $v$ is $\order{M/4\pi}$, then $g_+/g$ could be 
anywhere between zero and infinity, and there would be no remnants of the 
underlying $SU(2)$ symmetry.   

This lesson can be generalized. In general $g_+$ differs from $g$ as
\beq
   g_+ = \alpha g ~,
\eql{g-alpha-rel}
\eeq
where the factor $\alpha$ includes contributions from all the 
operators that can mix with $W_\mu^a J^{a\mu}$.  The relation $\alpha \simeq 1$ 
holds as long as $v \ll M/4\pi$, but for $v \sim M/4\pi$,
all those operators would contribute to $\alpha$ equally in magnitude, and 
consequently $\alpha$ could be anywhere between zero and infinity.   

Now, let us go back to the case of our interest and take $\Lg$ into account.
Let us write the doublet $\psi$ as
\beq
   \psi = \colmvec{ \chit \cr \chi } ~,
\eeq
and, for definiteness, take $\chit$ to be heavier than $\chi$ 
with the mass difference larger than $\mW$ so that $\chit$ can decay into 
$\chi$ and $W^+$.  Clearly, this is the analog of a sparticle decaying
into its SM partner and a gravitino.  

Once we have seen a toy gravitino produced via this decay, $g_+$ must be 
nonzero.  In the rest frame of the decaying $\chit$, this decay is caused by
the operator
\beq
  {\cal H}_{\rm int} 
     \supset g_+ \, W^+_{-{\bf p}} \,\chi_{\bf p} \,\chit_{\bf 0} 
  ~,
\eeq
where the irrelevant indices, bars and daggers are suppressed, while the
important quantity here is $\abs{\bf p} = \sqrt{E_\chi^2 - m_\chi^2}$ where 
$E_\chi$ is the energy of the outgoing $\chi$ given by
\beq
  E_\chi = \frac{m_\chit^2 + m_\chi^2 - \mW^2}{2m_\chit}  ~.
\eql{E_chi}  
\eeq

\begin{figure}
\includegraphics{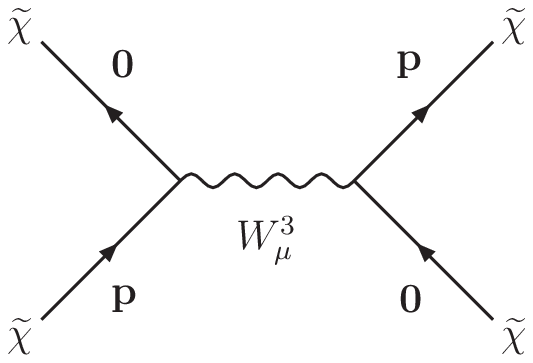}
\caption{The $\chit_{\bf p} \chit_{\bf 0} \to \chit_{\bf 0} \chit_{\bf p}$ 
scattering via the $t$-channel $W^3_\mu$ exchange.}
\label{fig:chichi}
\end{figure}

Now, if $g$ is also nonzero, there would also be a term  
\beq
  {\cal H}_{\rm int} 
     \supset g \, W^3_{-{\bf p}} \,\chit_{\bf p} \,\chit_{\bf 0}  ~,
\eql{bad-op}
\eeq
with the {\it same} ${\bf p}$.  The problem is that, at the second order in 
perturbation theory, this operator could cause the process 
$\chit_{\bf p} \,\chit_{\bf 0} \to \chit_{\bf 0} \,\chit_{\bf p}$ 
via the $t$-channel $W^3_\mu$ exchange (FIG.~\ref{fig:chichi}).  Note that
the momentum transfer $Q^2$ mediated by the $W^3_\mu$ is given 
by
\beq
  Q^2 &=& -\left( \sqrt{{\bf p}^2 + m_\chit^2} - m_\chit \right)^2 
           + {\bf p}^2  \nn\\
      &=& 2 m_\chit^2  \left( \sqrt{1+\frac{{\bf p}^2}{m_\chit^2}} 
           - 1 \right) ~.
\eeq
From $\abs{\bf p} = \sqrt{E_\chi^2 - m_\chi^2}$ and \eq{E_chi}, we see that
for generic $m_\chi$ and $\mW$, we have $\abs{Q} \sim m_\chit > \mW \gg \Lg$,
which is a {\it hard} momentum transfer.%
\footnote{The exception occurs in `highly degenerate' cases:~(a) either 
one of $\chi$ and $W^+$ is much heavier than the other and almost degenerate 
with $\chit$ so that $\abs{\bf p} \ll \Lg$ or (b) $m_\chi$ and $\mW$ are of 
the same order but they add up to nearly $m_\chit$ so that 
$\abs{\bf p} \ll \Lg$.  Although these case are logically possible, it looks 
too coincidental, so we will not pursue this caveat any further.}
However, as discussed in 
Sec.~\ref{sec:SGET}, a graviton cannot be exchanged to mediate such a 
large momentum transfer within SGET.  Therefore, the operator \eq{bad-op}
should not be present in the effective theory.

Therefore, to decouple the operator \eq{bad-op}, we must take the limit 
$g \to 0$ while keeping $g_+$ fixed to a finite value.  Then the relation
\eq{g-alpha-rel} requires $\alpha \to \infty$, which, however, is possible 
only if $v \sim M/4\pi$, as noted before.  In this limit, all 
the higher-dimensional operators that can contribute to $g_+$ do contribute 
equally in magnitude, while all the other interactions have the full 
NDA strength.  Furthermore, having taken this limit, we have decoupled the 
{\it soft} $W^3_\mu$ as well, so we have to couple it back to the theory.  
This can be easily done by using the 
$U(1)$ invariance, but now the coupling of this $U(1)$---let us call it 
$g_{\rm soft}$---is completely arbitrary, with no relation to $g_+$!    

Therefore, although we cannot perform any quantitatively reliable analysis 
beyond estimates%
\footnote{We have also neglected the effects of running.}
due to $v \sim M/4\pi$, this is good enough to give us the following 
qualitative understanding of what $W^+_\mu$ is like.  First, its coupling to the
$SU(2)$ current, $g_+$, is not related at all to the coupling of the soft 
$W^3_\mu$ to the $U(1)$ current, $g_{\rm soft}$.  Second, it has all kinds of
additional interactions, all with the full NDA strength.  Because of these
two features, $W^+_\mu$ should be viewed just as a random spin-1 resonance, 
rather than the `$SU(2)$-partner' of $W^3_\mu$.   

Recalling the dictionary of our analogy, translating this gauge-theory lesson 
back to gravity is straightforward.  (The only slight mismatch in the 
dictionary, which is not at all essential for us, appears in the $4\pi$ counting 
for broken SUSY, where the relation $v \sim M/4\pi$ should be translated as 
$F \sim M^2/4\pi$ where $F$ is the decay constant of the goldstino, or the 
square of the SUSY breaking scale \cite{Luty:1997fk}.)  Therefore, we have 
found  
\begin{itemize}
 \item{If $m_{3/2} \ll \Lg$, it is {\it consistent} for the gravitino to be 
  just `canonical', with all the properties we expect from the standard 
  supergravity, except for the fact that the gravitino---like the 
  graviton---is not an elementary degree of freedom at short distances.  
  In other words, as long as we avoid processes where a gravitino receives or 
  mediates a large momentum transfer, the gravitino can behave normally.}
 \item{If $m_{3/2} \gg \Lg$, this `gravitino' is not really a gravitino,
  because the coupling of this `gravitino' to a SM particle and its 
  superpartner can have any value, with no relation to the `canonical' 
  strength, and we also expect this `gravitino' to have a whole series of other 
  couplings, all equally important with the full NDA strength.  In short, it 
  behaves just like a random spin-$3/2$ resonance with no relation to the 
  gravity sector.} 
\end{itemize} 
Hereafter, to distinguish these cases, we will use the term 
{\it gravitino} only to refer to the first case, while we will call the 
second case {\it pseudo-gravitino.}  

We postpone the issue of experimentally distinguishing a gravitino from a 
pseudo-gravitino until Sec.~\ref{sec:detection}.  At this point, let us just 
assume that the distinction can be made.  Then, we have found the 
{\it model-independent} relation between the gravitino mass and the composite
gravity scale:
\beq
  m_{3/2} \lsim \Lg ~.
\eql{theprediction}
\eeq
By definition, gravity is described by GR at distances longer than $\Lg^{-1}$,
because GR is the only consistent theory once we have a graviton coupled to
matter described by a local relativistic quantum field theory 
\cite{Weinberg:1965rz}.  (Note that we could not have said this if we had not 
restricted $\Lg$ below $\tev$ which assures the matter sector is `normal'.)  
Therefore, the relation \eq{theprediction} 
means that {\it the existence of a gravitino guarantees that GR is correct 
at least down to its Compton wavelength!}  Hence, this is a short-distance 
test of GR, which in turn places a lower bound on $\Lg$.      
On the other hand, the relation \eq{theprediction} implies that if we find 
composite gravity first at some $\Lg$, then we will not discover a gravitino
above the scale $\Lg$---at best we may just see a pseudo-gravitino which is 
nothing but a random spin-3/2 state.

\section{Discussions}
\label{sec:discuss}

\subsection{Should a Gravitino Exist?}
\label{sec:existence}
The quick answer is, we don't know.  There is no strong argument indicating 
whether it should or shouldn't.  We will present below several arguments, 
not to answer this question but to shed different light and gain more 
insights on the result of Sec.~\ref{sec:SusySGET}.   

Imagine a huge hierarchy between the SUSY breaking scale $\sqrt{F}$ and $\Lg$, 
as $\sqrt{F} \gg \Lg$.  Above $\Lg$, the gravity sector is described by some 
exotic degrees of freedom---which may not even be field-theoretic---with no 
gravitons.  Here, there is no point of asking what the superpartner of the 
{\it graviton} is, because the graviton is not even in the theory.  
When we go below $\Lg$, the graviton emerges, but we do not expect that a 
gravitino appears there, because from the usual effective-field-theoretic 
viewpoint, the dynamics at $\Lg$ that generates the graviton should not 
`know' about SUSY which is broken \textit{way} above $\Lg$.  

This argument is too naive, however.  As we will argue below, not only is it 
possible that a 
pseudo-gravitino may exist, but also even an honest gravitino with all the 
(approximately)
canonical properties may exist!  Consider a supersymmetric $SU(3)$ gauge 
theory with two flavors with no superpotentials, and suppose that SUSY is 
broken with the soft masses much larger than $\Lambda_{\rm QCD}$.  For 
simplicity and definiteness, also assume that the squark masses are all 
degenerate, respecting the flavor symmetry, and that the gluino is much 
heavier than the squarks.  This theory possesses an $R$ parity under which 
all the quarks and gluon are even while all the squarks and gluino are odd.  
Hence, the squarks are stable, and there are stable fermionic meson-like bound 
states (`mesinos') with one quark and one anti-squark.%
\footnote{We are assuming that these mesinos are the lightest among the 
hadrons containing superparticles.}

So, apparently, the mesons have superpartners, the mesinos.  But look at other 
particles; for example, there is no `sproton' or `sneutron', because they 
would decay too quickly to form a bound state.  In fact, most particles lack 
their superpartner, so the interactions between the meson-mesino sector and 
the rest are completely non-supersymmetric.  Therefore, if these 
non-supersymmetric couplings are significant, there is no sense in which the
mesinos are the superpartners of the mesons, except for the quantum numbers.  
In other words, the mesinos in this case are just analogous to our 
pseudo-gravitino.

However, there is also a logical possibility that the couplings between the 
meson-mesino sector and the rest are sufficiently small for some reason. 
Then, it is at least {\it consistent} for the mesinos to retain the properties 
expected from supersymmetry.%
\footnote{Note that this is exactly what is happening in typical weak-scale 
SUSY models in which the visible-sector interactions at the weak scale are 
taken to be (approximately) supersymmetric, even though the actual SUSY 
breaking scale is often as high as $10^{11}\,\gev$. This is consistent because 
the interaction that transmits SUSY breaking to the visible sector is assumed 
to be sufficiently feeble.}
A similar situation could happen to a gravitino.  For example, if $\Lg$ is, say,
$10^{-2}~\ev$, then it could be perfectly consistent for a gravitino with, 
say, $m_{3/2} = 10^{-3}~\ev$ to carry all the (approximately) canonical 
couplings we expect from supergravity---as long as the gravitino does not 
receive or mediate a momentum transfer larger than $\Lg$---even though 
$\sqrt{F}$ here would be $\sim \tev$ which is way above $\Lg$.
To sum up, from the standard effective-field-theoretic view, there seems no 
preference among `nothing', `a pseudo-gravitino', and `a (real) gravitino'.     

To gain more insight, let us consider the limit in the opposite order.  This 
time we start with a finite $\Lg$ but no SUSY breaking ($F=0$).%
\footnote{An extreme but trivial limit of this case is to take 
$\Lg \gsim \Mpl$, \ie, the limit of an elementary graviton.  Note that for any 
$\sqrt{F} \lsim \Mpl$, the gravitino is a normal gravitino, and the inequality 
\eq{theprediction} is trivially satisfied since 
$m_{3/2} \sim F/\Mpl \lsim \Mpl \lsim \Lg$.}
So we start 
with a degenerate pair of massless graviton and gravitino.  This gravitino 
is of course exactly what we expect from supergravity, as long as we avoid 
momentum transfers larger than $\Lg$.  As we raise $F$, the gravitino mass 
goes up according to the usual relation $m_{3/2} \sim  F/\Mpl$, as long as 
$m_{3/2}  \ll \Lg$.  If we keep raising $F$, $m_{3/2}$ eventually hits
$\Lg$, beyond which the gravitino may start looking strange.  (The result of
Sec.~\ref{sec:SusySGET} says it {\it will} start looking strange, but here
let us pretend that we did not know Sec.~\ref{sec:SusySGET}.)  Then, in 
particular we no longer know how $m_{3/2}$ should vary as a function of $F$. 
(We will come back to this issue in detail in Sec.~\ref{sec:relation}.)  
Here, let us suppose that it still keeps going up, although not 
necessarily obeying the usual linear relation.  Will this `gravitino' 
eventually disappear?  Note that it will disappear from SGET if its 
lifetime becomes shorter than $\Lg^{-1}$.  Naively, we expect that the 
lifetime should be quite long because the coupling $1/\Mpl$ is extremely weak, 
so it would stay in the effective theory even if $m_{3/2}$ is as high as the 
weak scale.  But this `gravitino' may have unusual interactions, and there 
are probably many new states around $E \sim \Lg$ into which the `gravitino' 
could decay.  So the lifetime may or may not be quick enough for the 
`gravitino' to disappear from SGET.  We need the underlying theory to see 
which way it goes.   

Finally, it is also conceivable that $m_{3/2}$ `saturates' at $\Lg$ as we raise 
$F$.  We would expect this if there is an exotic state at $E\sim\Lg$ 
which can mix with the gravitino. Then, by the `no-level-crossing' theorem, 
$m_{3/2}$ cannot go up any further, and the `gravitino' becomes a mixture of 
the original gravitino and this exotic state.  Therefore, in this
case, we expect a pseudo-gravitino with $m_{3/2} \sim \Lg$.

To summarize, qualitative arguments seems completely inconclusive about 
the nature and fate of a gravitino.  The result of 
Sec.~\ref{sec:SusySGET} is therefore quite nontrivial.

\subsection{Relation of $m_{3/2}$ to SUSY Breaking Scale}
\label{sec:relation}
Here, we comment on the validity of the famous relation between 
the gravitino mass and the SUSY breaking scale:
\beq
   m_{3/2} = \frac{F}{\sqrt{3} \Mpl}  ~.
\eql{famous-rel}
\eeq
In the pseudo-gravitino case ($m_{3/2} \gg \Lg$), this usual relation
has no reason to be true.  Clearly, we cannot use the supergravity formalism 
to derive it, because supergravity contains general relativity which is not 
applicable for $E \gg \Lg$ in our scenario.  But more fundamentally, recall 
that this relation is just a consequence of the equivalence between the 
goldstino and the longitudinal component of the gravitino at high energies 
($E_{3/2} \gg m_{3/2}$).  Usually, we derive the relation by demanding that 
the amplitude of exchanging a gravitino between two supersymmetry currents 
be equal to that of exchanging a goldstino, in the global SUSY limit 
($\Mpl \to \infty$) for $E_{3/2} \gg m_{3/2}$.  However, in the 
pseudo-gravitino case, it has a different coupling to the supersymmetry 
current as well as a host of additional interactions.  Hence, the formula
\eq{famous-rel} does not hold for a pseudo-gravitino.  In other words,
since the pseudo-gravitino does not eat the goldstino by exactly the right
amount, the SUSY currents must exchange something else to match the 
goldstino-exchange amplitude.  But this `something else' must be among the 
new exotic states in the full theory of gravity, which we have no idea about.  
(If we had the underlying theory, we could subtract the exotic contribution 
from the amplitude and figure out how the formula \eq{famous-rel} should get 
modified.)

On the other hand, for $m_{3/2} \ll \Lg$, we can apply the derivation 
for $m_{3/2} \ll E_{3/2} \ll \Lg$, and obtain the usual relation 
\eq{famous-rel}, assuming that the gravitino has the standard $1/\Mpl$ 
coupling to the SUSY current, which is at least a consistent thing to do as 
we discussed in Sec.~\ref{sec:SusySGET}.

\subsection{Theoretical Tests}
\label{sec:theo-tests}
It is certainly desirable to confirm the result of Sec.~\ref{sec:SusySGET}
by a theoretical argument that has a firmer foundation.  Recall the 
concrete example of composite gauge bosons mentioned in Sec.~\ref{sec:intro}:
the $SU(N)$ supersymmetric QCD with $F$ flavors, where $N+1<F<3N/2$.  
Below the $\Lambda_{\rm QCD}$ of the $SU(N)$, this theory is described in 
terms of an IR-free $SU(F-N)$ gauge theory whose gauge bosons are composites 
of the original degrees of freedom \cite{Seiberg:1994pq}.  

Now let us deform the theory such that the low-energy gauge group $SU(F-N)$ 
gets spontaneously broken down to $SU(M)$ where $M<F-N$.  If we apply the
argument of Sec.~\ref{sec:SusySGET} to this theory, we predict that
the massive $W$ bosons, with all the `normal' couplings retained, cannot be 
heavier than $\Lambda_{\rm QCD}$.  $W$ bosons heavier than $\Lambda_{\rm QCD}$ 
may exist but they should behave like random spin-1 resonances, rather than 
as the `$SU(F-N)$-partners' of the $SU(M)$ gauge bosons.  While it sounds 
plausible, the currently available theoretical wisdoms are not powerful 
enough to definitively confirm the statement.

This SUSY QCD example also illustrates how extremely nontrivial it is 
to have a composite graviton coupled to elementary matter particles.  
In the case of the SUSY QCD model, this corresponds to the composite $SU(F-N)$ 
gauge bosons coupled to elementary quarks that are point-like even far above
$\Lambda_{\rm QCD}$!  This is clearly a very difficult, if possible, thing
to do.  In the AdS composite graviton model of Ref.~\cite{Gherghetta:2005se}, 
the graviton wavefunction is highly peaked toward the IR brane, but there is 
an exponentially suppressed tail overlapping the UV brane where the SM fields 
live, which can be thought of as an explanation for the weakness of gravity.  
Adding supersymmetry to their setup to study the gravitino properties is
saved for future work.

\section{Precision Gravitino Study and Probing $\Lg$ in Colliders}
\label{sec:detection}
Clearly, the most important quantity in any composite graviton scenario is 
the scale $\Lg$.  As we mentioned already in Sec.~\ref{sec:SusySGET}, 
in order to probe the scale $\Lg$, it is crucial to experimentally 
distinguish a gravitino from a pseudo-gravitino.  

Unfortunately, if the results of such `precision gravitino study' turn out 
that what we have seen is actually a pseudo-gravitino, this will not be a 
sufficient evidence that gravity is modified at short distances.  For 
example, a pseudo-gravitino is also present in a scenario where supersymmetry 
is not a fundamental symmetry at high energies but merely an (approximate) 
{\it accidental} global symmetry of the matter sector at low energies
\cite{Goh:2003yr}.  In this scenario, the gravity sector is just the 
conventional GR (with no supersymmetry).  Therefore, for a pseudo-gravitino, 
we need the underlying theory to derive more specific predictions to be 
tested. 

On the other hand, if we can convince ourselves that it is not a 
pseudo-gravitino, then we can put a {\it model-independent} lower bound 
on $\Lg$, as $\Lg \gsim m_{3/2}$!  Interestingly, as we will see shortly,
in precisely the regime 
that the direct gravity measurement between test masses is impossible, the 
measurement of $m_{3/2}$ becomes possible, so the precision gravitino study 
can potentially exclude composite graviton scenarios dramatically at very
short distances.   

Since it is impossible to see a gravitino $\psi_{3/2}$ directly, the only hope 
to learn 
something about it lies in the case where both $\Xt$ and $X$ can be precisely
studied in the decay $\Xt \to X + \psi_{3/2}$.  This means that the decay 
must be sufficiently slow and that $\Xt$ and $X$ both must be visible.  This 
will indeed be realized if the $\Xt$ is the next-to-lightest supersymmetric 
particle (NLSP) (the lightest (LSP) being the gravitino) and is electrically 
charged and/or strongly-interacting.  In such a case, due to the very weak 
coupling of $\Xt$ to the gravitino, there will be a long, highly visible 
track of the NLSP inside a collider detector before it decays 
\cite{Drees:1990yw}, unless $\psi_{3/2}$ is too light.  It is even possible 
that the NLSP stops in the detector if it is strongly interacting or produced 
sufficiently slow.  In such circumstances, the momenta and energies of the 
NLSP and its SM partner as well as the NLSP lifetime should be measurable, 
which in turn allows us to deduce the mass and the coupling of the gravitino 
to see whether it is a pseudo-gravitino or not. 

This `gravitino LSP with charged NLSP' scenario has already been a 
great interest in SUSY phenomenology, especially in the context of 
gauge-mediated SUSY models where the gravitino is the LSP and $\Xt$ is often 
a scalar tau lepton \cite{Dimopoulos:1996vz, Buchmuller:2004rq, 
Hamaguchi:2004df}.  Note that once $X$ and $\Xt$ have been observed, the 
gravitino mass can be simply determined from rewriting \eq{E_chi} :
\beq
   m_{3/2} = \left( m_X^2 + m_\Xt^2 - 2 m_\Xt E_X \right)^{1/2} ~,
\eql{kinematics}
\eeq
where $E_X$ is the energy of the $X$ measured in the rest frame of the $\Xt$.
If $\Xt$ stops inside a detector, $E_X$ can be directly measured.  
Even if it does not, since both the $X$ and $\Xt$ are highly visible in the 
detector, the measurement of their energies and the relative angle (the `kink'
in the track) can determine $E_X$.    

On the other hand, the measurement of the $\Xt$ lifetime gives us the 
gravitino's coupling.  If what we are seeing is not a pseudo-gravitino but
is a real one, then the coupling should go as $1/\Mpl$ times the polarization
factor $E_{3/2}/m_{3/2}$ for the helicity-$\pm 1/2$ components, so the rate 
is given by
\beq
\eql {NLSP-rate}
  \Gamma_\Xt
        &=&    \frac{m_\Xt^5}{48\pi \Mpl^2 \, m_{3/2}^2}  \nn\\
     &\approx& (20~\micr)^{-1} 
                \left( \frac{\ev}{m_{3/2}} \right)^2
                \left( \frac{m_\Xt}{100\,\gev} \right)^5  \\
     &\approx& (20~{\rm hours})^{-1} 
                \left( \frac{\gev}{m_{3/2}} \right)^2
                \left( \frac{m_\Xt}{100\,\gev} \right)^5  \nn
\eeq
where we have dropped $m_X$ and $m_{3/2}$ for simplicity. (The 
helicity-$\pm 3/2$ components have no $E_{3/2}/m_{3/2}$ enhancement and thus 
have been neglected.) 
The consistency of $m_{3/2}$ determined from this formula with the value 
extracted from pure kinematics \eq{kinematics} will be an almost convincing 
evidence that the gravitino is not a pseudo, because it would be such a 
coincidence if the pseudo-gravitino coupling, which could be any size, just 
happened to be $1/\Mpl$.%
\footnote{Note that this agreement between the two measurements
of $m_{3/2}$ is equivalent to checking if the gravitino has really eaten the
goldstino as it should if it is not a pseudo.}
Ref.~\cite{Buchmuller:2004rq} proposes to go even further, to test the 
gravitino's {\it spin} by using the angular distribution in the 3-body decay 
$\tilde{\tau} \to \tau + \gamma + \psi_{3/2}$.                           

Now, it is probably extremely hard to directly measure the gravitational force 
between test masses for distances smaller than the micron scale which 
would correspond to $\Lg \lsim 10^{-1}\,\ev$.  Let us see whether the 
precision gravitino study can be used to place a bound on $\Lg$ beyond this 
limitation.  Taking $m_\Xt = 100~\gev$, the rate \eq{NLSP-rate} tells us that
for $m_{3/2} = 10^{-1}\,\ev$, the NLSP will decay within $\order{1}\,\micr$, 
since the relativistic $\gamma$ factor for the NLSP cannot be larger than
$\order{10}$ in a $\tev$-scale collider.  This is unfortunately too short 
to be seen.  Demanding that the NLSP must fly at least a few 
$100\,\micr$ to be clearly observed by a micro vertex detector, we need 
$m_{3/2}$ to be at least a few $\ev$.  However, for such low values for 
$m_{3/2}$, the formula \eq{kinematics} requires $m_X$, $m_\Xt$ and $E_X$ to be 
measured with unrealistically high precision.  The problem is, to determine 
a small $m_{3/2}$ from \eq{kinematics}, we have to nearly cancel two large
terms and take the square-root.  Therefore, the lowest possible value for 
$\Lg$ that can be probed is actually limited by the accuracy in measuring 
these parameters rather than the minimal NLSP flight length that a detector 
can resolve.  For example, if we are anticipating 
$m_{3/2}$ of order $1~\gev$ and if we are content with determining 
$m_{3/2}$ only up to a factor of a few, then for $m_\Xt \approx 100~\gev$ 
(neglecting $m_X$ for simplicity), we would need to measure $m_\Xt$ and $E_X$ 
with the accuracy of $\pm 10~\Mev$.  Therefore, measuring $m_{3/2}$ of 
$\order{1}\,\gev$ event-by-event is unrealistic, so it must be done 
statistically.  Taking the uncertainty in the individual $E_X$ 
measurement to be $\order{1}\,\gev$, we need to observe $\order{10^4}$ NLSP 
decays to have enough statistics for $m_\Xt \approx 100~\gev$ and 
$m_{3/2} \sim \gev$.  

Also, note that for $m_{3/2} \sim \gev$, the $\Xt$ lifetime is about 
a few hours to a week, so the NLSPs must be collected and stored to
do the measurement.  Such a possibility for $\Xt = \tilde{\tau}$ has been 
extensively studied in Refs.~\cite{Hamaguchi:2004df}, and the bottom line 
is that collecting $\order{10^4}$ or even $\order{10^5}$ NLSPs and observing 
their decays should be possible in the LHC and/or the ILC, although the
prospect depends on other SUSY parameters.

Those analyses also conclude that we may be able to go up to $m_{3/2}$ 
of $\order{100}\,\gev$.  Therefore, it is not too optimistic to expect that
precision gravitino study may be able to probe the scale $\Lg$ between 
$\gev$ and $100~\gev$.  While this is still quite challenging 
(and we also have to be lucky with the SUSY spectrum), note that this is a 
regime where direct measurement of gravitational forces is absolutely 
impossible, so precision gravitino study is the only available probe for 
composite gravity.

\section{Conclusions}
\label{sec:conc}
In this paper, we have considered `composite gravity', namely, the 
possibility that the graviton is not an elementary propagating degree of 
freedom at distances shorter than $\Lg^{-1}$.  We pointed out that such
a scenario is not necessarily forbidden by the Weinberg-Witten theorem.
Another important assumption we made is that the matter sector is completely 
described by a local quantum field theory, which is true for $\Lg$ between 
the current experimental limit $\sim 10^{-3}\,\ev$, and $\sim \tev$ or whatever 
cutoff for the matter sector.  To perform a model-independent, 
effective-field-theoretic analysis, it is necessary to reconcile 
`elementary matter with a high cutoff' and `composite gravity
with a low cutoff', and for this purpose we have utilized soft graviton 
effective theory (SGET) by Sundrum.

In general, the only way to place a lower limit on the scale $\Lg$ is by 
a null result in experiments seeking a deviation from the standard $1/r^2$ 
law between macroscopic test masses.  This method becomes increasingly 
difficult as the distance gets reduced.  Therefore, it is desirable to have 
an alternative probe.  The problem is, however, that in general there is 
nothing related to gravity except the graviton itself, so there is no other 
way to probe $\Lg$ without using gravity.

However, we noted that if there is an underlying supersymmetry, it may
lead to the existence of a gravitino, which is related to gravity but easier
to observe than the graviton.  Applying the SGET framework to the gravitino,
we have shown the relation, $\Lg \gsim m_{3/2}$, \ie, the {\it graviton} 
remains an elementary degree of freedom at least down to the {\it gravitino's} 
Compton wavelength.  In other words, we can use a gravitino to test general
relativity at short distances---once we see a gravitino, we know that GR is
correct at least up to $m_{3/2}$, which in turn places a lower bound on $\Lg$!  
This can have a significant impact on the possibility of composite graviton as
a solution to the cosmological constant problem.  For example, if we find 
$m_{3/2}$ to be, say, $1\>\gev$, the door will be completely shut. 

On the other hand, if we first find gravity compositeness and measure $\Lg$, 
then our inequality says that we will not discover a gravitino above the scale 
$\Lg$---at best we may just see some random spin-$3/2$ fermion with completely 
random couplings, nothing to do with gravity.  
    
To utilize this inequality to place a limit on $\Lg$, it is crucial to
experimentally convince ourselves that what we are looking at is really
a gravitino, rather than a random spin-$3/2$ fermion.  In the future 
colliders such as the LHC and ILC, the prospect of being able to do so 
seems quite bright for the range $\gev \lsim  m_{3/2} \lsim 100~\gev$,
corresponding to probing $\Lg$ in the range between 
$10^{-14}~{\rm cm}$ and $10^{-16}~{\rm cm}$.  Therefore, precision 
gravitino study can indeed be an alternative model-independent probe for 
$\Lg$ or a test of general relativity, in a regime where direct measurement 
of gravitational force is absolutely impossible.

\begin{acknowledgments}
I thank Zackaria Chacko, Markus Luty, Yasunori Nomura, Matt Schwartz, 
Raman Sundrum, and Mithat \"Unsal for discussions and comments on the 
manuscript.  I also thank Spencer Chang, Hitoshi Murayama, and Michael Peskin 
for conversations.  In addition, I thank the Aspen Center for Physics where
a portion of this work was conducted.  This work was supported by DOE grant 
DE-FG02-91ER40676.   
\end{acknowledgments}


\begin{thebibliography}{99}

\bibitem{Hoyle:2004cw}
C.~D.~Hoyle, D.~J.~Kapner, B.~R.~Heckel, E.~G.~Adelberger, J.~H.~Gundlach, U.~Schmidt and H.~E.~Swanson,
Phys.\ Rev.\ D {\bf 70}, 042004 (2004)
[arXiv:hep-ph/0405262].


\bibitem{Long:2003dx}
J.~C.~Long, H.~W.~Chan, A.~B.~Churnside, E.~A.~Gulbis, M.~C.~M.~Varney and J.~C.~Price,
Nature {\bf 421}, 922 (2003).


\bibitem{Smullin:2005iv}
S.~J.~Smullin, A.~A.~Geraci, D.~M.~Weld, J.~Chiaverini, S.~Holmes and A.~Kapitulnik,
arXiv:hep-ph/0508204.


\bibitem{Adelberger:2003zx}
E.~G.~Adelberger, B.~R.~Heckel and A.~E.~Nelson,
Ann.\ Rev.\ Nucl.\ Part.\ Sci.\  {\bf 53}, 77 (2003)
[arXiv:hep-ph/0307284].


\bibitem{:2005em}
The ALEPH, DELPHI, L3, OPAL, SLD Collaborations,
the LEP Electroweak Working Group, and
the SLD Electroweak and Heavy Flavour Groups
arXiv:hep-ex/0509008;\\
S.~Eidelman {\it et al.}  [Particle Data Group],
Phys.\ Lett.\ B {\bf 592} (2004) 1;


\bibitem{Arkani-Hamed:1998rs}
N.~Arkani-Hamed, S.~Dimopoulos and G.~R.~Dvali,
Phys.\ Lett.\ B {\bf 429}, 263 (1998)
[arXiv:hep-ph/9803315].


\bibitem{Sundrum:1997js}
R.~Sundrum,
JHEP {\bf 9907}, 001 (1999)
[arXiv:hep-ph/9708329].


\bibitem{Moffat:2001jf}
J.~W.~Moffat,
arXiv:hep-ph/0102088.


\bibitem{Sundrum:2003tb}
R.~Sundrum,
Nucl.\ Phys.\ B {\bf 690}, 302 (2004)
[arXiv:hep-th/0310251].


\bibitem{Unruh:1980cg}
W.~G.~Unruh,
Phys.\ Rev.\ Lett.\  {\bf 46}, 1351 (1981);\\
M.~Visser,
arXiv:gr-qc/9311028;\\
L.~J.~Garay, J.~R.~Anglin, J.~I.~Cirac and P.~Zoller,
Phys.\ Rev.\ Lett.\  {\bf 85}, 4643 (2000)
[arXiv:gr-qc/0002015];\\
G.~Chapline, E.~Hohlfeld, R.~B.~Laughlin and D.~I.~Santiago,
Int.\ J.\ Mod.\ Phys.\ A {\bf 18}, 3587 (2003)
[arXiv:gr-qc/0012094].


\bibitem{Weinberg:1980kq}
S.~Weinberg and E.~Witten,
Phys.\ Lett.\ B {\bf 96}, 59 (1980).


\bibitem{Seiberg:1994pq}
N.~Seiberg,
Nucl.\ Phys.\ B {\bf 435}, 129 (1995)
[arXiv:hep-th/9411149].


\bibitem{Gherghetta:2005se}
T.~Gherghetta, M.~Peloso and E.~Poppitz,
arXiv:hep-th/0507245.


\bibitem{Perlmutter:1998np}
S.~Perlmutter {\it et al.}  [Supernova Cosmology Project Collaboration],
Astrophys.\ J.\  {\bf 517}, 565 (1999)
[arXiv:astro-ph/9812133];\\
A.~G.~Riess {\it et al.}  [Supernova Search Team Collaboration],
Astrophys.\ J.\  {\bf 607}, 665 (2004)
[arXiv:astro-ph/0402512].


\bibitem{Spergel:2003cb}
D.~N.~Spergel {\it et al.}  [WMAP Collaboration],
Astrophys.\ J.\ Suppl.\  {\bf 148}, 175 (2003)
[arXiv:astro-ph/0302209];\\
P.~de Bernardis {\it et al.}  [Boomerang Collaboration],
Nature {\bf 404}, 955 (2000)
[arXiv:astro-ph/0004404];\\
A.~Balbi {\it et al.},
Astrophys.\ J.\  {\bf 545}, L1 (2000)
[Erratum-ibid.\  {\bf 558}, L145 (2001)]
[arXiv:astro-ph/0005124].


\bibitem{Tegmark:2003ud}
M.~Tegmark {\it et al.}  [SDSS Collaboration],
Phys.\ Rev.\ D {\bf 69}, 103501 (2004)
[arXiv:astro-ph/0310723];\\
J.~A.~Peacock {\it et al.},
Nature {\bf 410} (2001) 169
[arXiv:astro-ph/0103143].


\bibitem{Kaplan:2005rr}
D.~E.~Kaplan and R.~Sundrum,
arXiv:hep-th/0505265.


\bibitem{Sundrum:2003jq}
R.~Sundrum,
Phys.\ Rev.\ D {\bf 69}, 044014 (2004)
[arXiv:hep-th/0306106].


\bibitem{Goldberger:2004jt}
W.~D.~Goldberger and I.~Z.~Rothstein,
arXiv:hep-th/0409156.


\bibitem{Donoghue:1995cz}
J.~F.~Donoghue,
arXiv:gr-qc/9512024;\\
C.~P.~Burgess,
Living Rev.\ Rel.\  {\bf 7}, 5 (2004)
[arXiv:gr-qc/0311082].


\bibitem{Georgi:1990um}
H.~Georgi,
Phys.\ Lett.\ B {\bf 240}, 447 (1990).


\bibitem{Kaplan:1995uv}
D.~B.~Kaplan,
arXiv:nucl-th/9506035.


\bibitem{Luke:1996hj}
M.~Luke and A.~V.~Manohar,
Phys.\ Rev.\ D {\bf 55}, 4129 (1997)
[arXiv:hep-ph/9610534].


\bibitem{Manohar:1983md}
A.~Manohar and H.~Georgi,
Nucl.\ Phys.\ B {\bf 234}, 189 (1984);\\
H.~Georgi,
``Weak Interactions And Modern Particle Theory,'';\\
H.~Georgi and L.~Randall,
Nucl.\ Phys.\ B {\bf 276}, 241 (1986).


\bibitem{Luty:1997fk}
M.~A.~Luty,
Phys.\ Rev.\ D {\bf 57}, 1531 (1998)
[arXiv:hep-ph/9706235];\\
A.~G.~Cohen, D.~B.~Kaplan and A.~E.~Nelson,
Phys.\ Lett.\ B {\bf 412}, 301 (1997)
[arXiv:hep-ph/9706275].


\bibitem{Weinberg:1965rz}
S.~Weinberg,
Phys.\ Rev.\  {\bf 138}, B988 (1965).


\bibitem{Goh:2003yr}
H.~S.~Goh, M.~A.~Luty and S.~P.~Ng,
JHEP {\bf 0501}, 040 (2005)
[arXiv:hep-th/0309103].


\bibitem{Drees:1990yw}
M.~Drees and X.~Tata,
Phys.\ Lett.\ B {\bf 252}, 695 (1990).


\bibitem{Dimopoulos:1996vz}
S.~Dimopoulos, M.~Dine, S.~Raby and S.~Thomas,
Phys.\ Rev.\ Lett.\  {\bf 76}, 3494 (1996)
[arXiv:hep-ph/9601367].


\bibitem{Buchmuller:2004rq}
W.~Buchmuller, K.~Hamaguchi, M.~Ratz and T.~Yanagida,
Phys.\ Lett.\ B {\bf 588}, 90 (2004)
[arXiv:hep-ph/0402179].


\bibitem{Hamaguchi:2004df}
K.~Hamaguchi, Y.~Kuno, T.~Nakaya and M.~M.~Nojiri,
Phys.\ Rev.\ D {\bf 70}, 115007 (2004)
[arXiv:hep-ph/0409248];
J.~L.~Feng and B.~T.~Smith,
Phys.\ Rev.\ D {\bf 71}, 015004 (2005)
[Erratum-ibid.\ D {\bf 71}, 0109904 (2005)]
[arXiv:hep-ph/0409278].

\end{thebibliography}
\end{document}